\documentclass[prb,reprint,superscriptaddress]{revtex4-2}
\pdfoutput=1
\usepackage{braket,dsfont,amsfonts,amssymb,bm,graphicx,float,amsmath,hyperref,amsthm}
\usepackage{mathrsfs, xcolor, soul}
\usepackage{orcidlink}

\newcommand{\rrangle}{\rangle\!\rangle}

\newcommand{\llpipe}{|}
\newcommand{\sket}[1]{\ensuremath{\llpipe\rrangle}}

\begin{document}
\title{Triggering boundary phase transitions through bulk measurements in two-dimensional cluster states}
\author{Yuchen Guo~\orcidlink{0000-0002-4901-2737}}
\affiliation{State Key Laboratory of Low Dimensional Quantum Physics and Department of Physics, Tsinghua University, Beijing 100084, China}
\author{Jian-Hao Zhang~\orcidlink{0000-0002-4455-0691}}
\affiliation{Department of Physics, The Pennsylvania State University, University Park, Pennsylvania 16802, USA}
\author{Zhen Bi~\orcidlink{0000-0003-0351-3963}}
\email{zjb5184@psu.edu}
\affiliation{Department of Physics, The Pennsylvania State University, University Park, Pennsylvania 16802, USA}
\author{Shuo Yang~\orcidlink{0000-0001-9733-8566}}
\email{shuoyang@tsinghua.edu.cn}
\affiliation{State Key Laboratory of Low Dimensional Quantum Physics and Department of Physics, Tsinghua University, Beijing 100084, China}
\affiliation{Frontier Science Center for Quantum Information, Beijing 100084, China}
\affiliation{Hefei National Laboratory, Hefei 230088, China}
\begin{abstract}
We investigate the phase diagram at the boundary of an infinite two-dimensional cluster state subject to bulk measurements using tensor network methods.
The state is subjected to uniform measurements $M = \cos{\theta}Z+\sin{\theta}X$ on the lower boundary qubits and in all bulk qubits.
Our results show that the boundary of the system exhibits volume-law entanglement at the measurement angle $\theta = \pi/2$ and area-law entanglement for any $\theta < \pi/2$. 
Within the area-law phase, a phase transition occurs at $\theta_c=1.371$. 
The phase with $\theta \in(\theta_c,\pi/2)$ is characterized by a noninjective matrix product state, which cannot be realized as the unique ground state of a one-dimensional local, gapped Hamiltonian.
Instead, it resembles a cat state with spontaneous symmetry breaking. 
These findings demonstrate that the phase diagram of the boundary of a two-dimensional system can be more intricate than that of a standard one-dimensional system. 
\end{abstract}
\maketitle

\section{Introduction}

There has been a growing interest in the measurement-induced phase transitions (MIPT) observed in random quantum circuits~\cite{Li2018, Skinner2019, Li2019} and tensor networks~\cite{Vasseur2019, Yang2022}. The steady states of these quantum evolutions can have different entanglement structures, such as area-law entanglement and volume=law entanglement, with respect to the measurement rate or other relevant parameters.
Numerical simulations have demonstrated~\cite{Li2018, Li2019} that these phase transitions exhibit universal critical scaling, some of which can be understood analytically through mapping into statistical models~\cite{Skinner2019, Vasseur2019, Jian2020, Bao2020}. 

These studies are also relevant to measurement-based quantum computation (MBQC)~\cite{Briegel2001, Raussendorf2001, Briegel2009}, where measurements on bulk qubits of a resource state can enable universal quantum computation at the boundary.
Ground states with symmetry-protected topological order (SPT)~\cite{Chen2010, Chen2011} can serve as resource states for MBQC in one dimension (1D)~\cite{Else2012, Stephen2017, Raussendorf2017} and two dimensions (2D)~\cite{Wei2017, Raussendorf2019, Stephen2019}.
Reference \cite{Liu2022} identified an entanglement phase transition at the boundary of the 2D cluster state between the area law and volume law when bulk states are measured randomly along $X$ or $Z$, which also implies possible MBQC with SPT phases~\cite{Raussendorf2019}.
However, the measurements are restricted to stabilizers since numerical simulations are strictly limited to small systems when adopting arbitrary measurements.
In addition, the randomness in the measurement configuration (both directions and outcomes) may obscure the detailed entanglement structure we observed and only differentiate between area law and volume law.
Therefore, one may fail to identify more phases in such systems as proposed in the main idea of MBQC.

In this paper, we explore using measurements beyond stabilizers to identify additional quantum phases in the boundary state.
By implementing uniform measurements for all qubits with postselection and adjusting the measurement basis, we eliminate randomness and reveal different phases.
Using the tensor network algorithm to directly simulate the thermodynamic limit, we identify three distinct quantum phases: a trivial phase with area-law entanglement, a nontrivial phase with area-law but a twofold degenerate entanglement spectrum, and a volume-law quantum phase.
Of particular interest is the intermediate phase. 
This phase exhibits a unique entanglement structure that cannot be realized as ground states of 1D local, gapped Hamiltonians but resembles a cat state of spontaneous symmetry breaking.
We demonstrate that this nontrivial phase can be detected by standard two-site correlation functions, such as $C_X(L)$ and $C_Z(L)$, and explore the properties of the corresponding phase transitions.

\begin{figure*}
    \centering
    \includegraphics[width=0.65\linewidth]{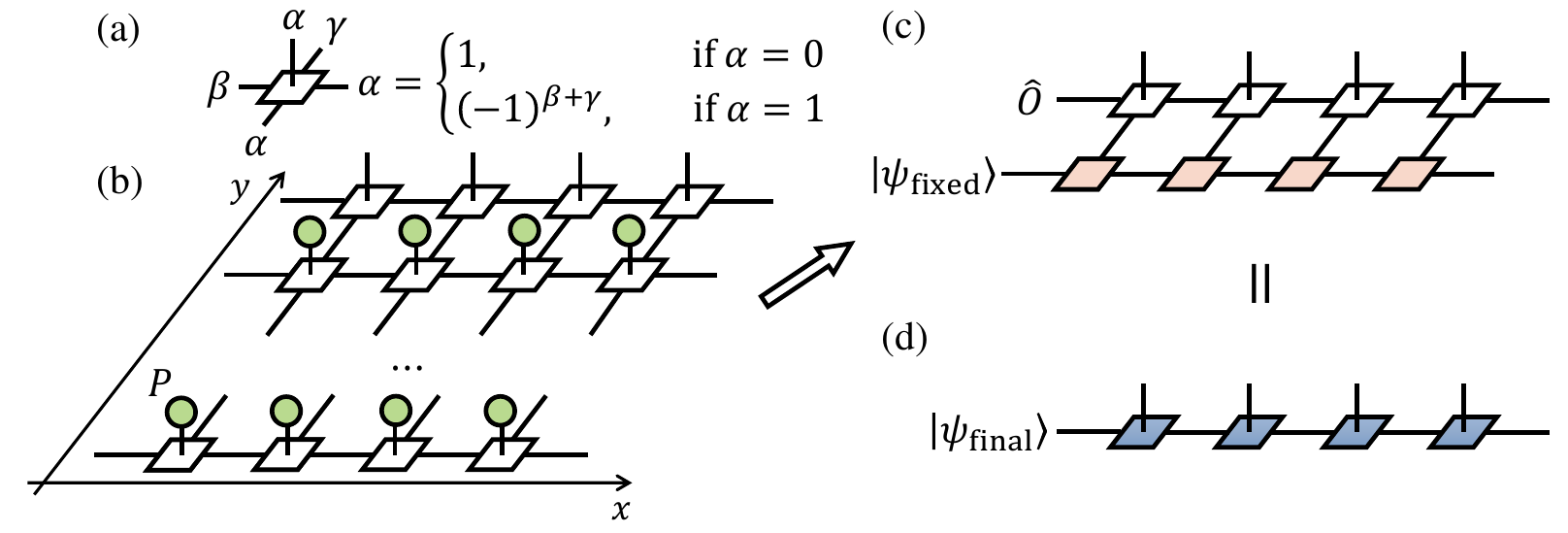}
    \caption{Settings of the problem.
    (a) PEPS construction for the cluster state with $D = 2$.
    (b) Measurement $P$ applied to bulk tensors.
    (c) The contraction of $L_y\rightarrow \infty$ layers results in $\ket{\psi_{\rm{fixed}}}$.
    (d) The final edge state $\ket{\psi_{\text{final}}}=\hat{O}\ket{\psi_{\text{fixed}}}$.}
    \label{Fig: settings}
\end{figure*}

\section{Measurements on 2D cluster state} 
On any graph with a spin-$1/2$ Hilbert space on each vertex, one can define a cluster state through a finite depth circuit
\begin{align}
    \ket{\Psi} = \prod_{l}{{\rm CZ}_l}\ket{+}^{\otimes N},\label{equ: cluster}
\end{align}
where $|+\rangle$ is the eigenstate of $X$ operator with +1 eigenvalues, and $\text{CZ}_{l}$ is the control-z gate $\mathrm{CZ}=\mathrm{diag}(1,1,1,-1)$ applied on each pair of spins on the $l$th link of the graph. 
The cluster state will be the ground state of the stabilizer Hamiltonian
\begin{align}
H= -\sum_{i}X_i\prod_{\langle i, j\rangle}{Z_j}.
\end{align}

In this work, we consider the cluster state on a square lattice which is infinite along the $x$ direction and has an open boundary condition (OBC) along the $y$ direction. 
We measure all the spins in the bulk of the square lattice and investigate how the resulting state on the boundary changes as we tune the bulk measurements. 
In particular, all measurements in the bulk are uniformly implemented by the projection operator $P=\frac{1}{2}(I+Z\cos{\theta}+X\sin{\theta}$), which is a projective measurement onto the $+1$ state of the operator $Z\cos{\theta}+X\sin{\theta}$. 
The measurement angle $\theta$ is the tuning parameter in our study. 

To address the problem, we adopt an infinite projected entangled pair state (iPEPS) representation for the cluster state.
The tensor in this iPEPS representation has bond dimension $D = 2$~\cite{Verstraete2008, Orus2014, Bridgeman2017} and the nonzero elements of the tensor are shown in Fig.~\ref{Fig: settings}(a).
Tensors after measurement can be obtained by contracting them with projectors $P$, as shown in Fig.~\ref{Fig: settings}(b).

For practical reasons, we consider a finite width of the system along the $y$ direction. 
For a given measurement angle $\theta$, we construct the upper boundary tensor without measurement ($D\times D\times D\times d_p$), the lower boundary tensor after measurement ($D\times D\times D$), and the bulk tensor after measurement ($D\times D\times D\times D$).
Therefore, we encounter a 1+1D dynamical problem on the virtual indices for the boundary state, where an initial infinite matrix product state (iMPS) whose physical indices are the virtual indices of the original boundary PEPS is evolved under multiple layers of infinite matrix product operators (iMPOs).
This problem can be simulated and analyzed with standard tensor network approaches~\cite{PerezGarcia2007, Verstraete2008, Orus2014, Bridgeman2017, Cirac2021}.

We focus on the convergence property for $L_y \rightarrow \infty $.
We begin with the lower boundary state (denoted as $\ket{\psi_{\text{init}}}$). 
Contracting one layer of bulk tensors is equivalent to implementing an iMPO, which we label as $\hat{H}$, on the iMPS.
In each iterative step, we contract the boundary tensors with one layer of bulk tensors and truncate them to a finite bond dimension $\chi$.
The resulting iMPS after implementing $L_y\rightarrow \infty$ layers of iMPO is denoted as the fixed-point iMPS $\ket{\psi_{\text{fixed}}}\propto \lim_{L_y\rightarrow \infty}\hat{H}^{L_y}\ket{\psi_{\text{init}}}$ shown in Fig.~\ref{Fig: settings}(c), which should be the eigenstate of $\hat{H}$ with the largest magnitude eigenvalue.
Therefore, we can also adopt the variational uniform matrix product state (VUMPS) method~\cite{Zauner2018, Vanderstraeten2019} to directly find the fixed point iMPS $\ket{\psi_{\text{fixed}}}$ of $\hat{H}$.
These two methods can produce consistent results in all the following numerical experiments.

After that, $\ket{\psi_{\text{fixed}}}$ is contracted with the upper boundary.
This contraction is equivalent to an iMPO with $D = 2$ applying a map from virtual indices to physical indices, i.e., $\ket{\psi_{\text{final}}}=\hat{O}\ket{\psi_{\text{fixed}}}$, as shown in Fig.~\ref{Fig: settings}(d).
It can be explicitly derived from the boundary PEPS that $\hat{O} = \prod_{i}{\text{CZ}_{i,i+1}}$.
In other words, $\hat{O}$ is just the product of local unitary (LU) transformations, i.e., $\ket{\psi_{\rm final}}\stackrel{\rm LU}{\sim}\ket{\psi_{\rm fixed}}$.
It does not affect the global entanglement structure, i.e., entanglement robust against renormalization group (RG) flow defined by a generalized local unitary (gLU) transformation~\cite{Chen2010, Chen2011, Zeng2019}.
In this sense, the global entanglement property of the final (upper) boundary iMPS is completely determined and characterized by the bulk tensors.

\section{Numerical simulations} 
\begin{figure*}
\includegraphics[width=0.99\linewidth]{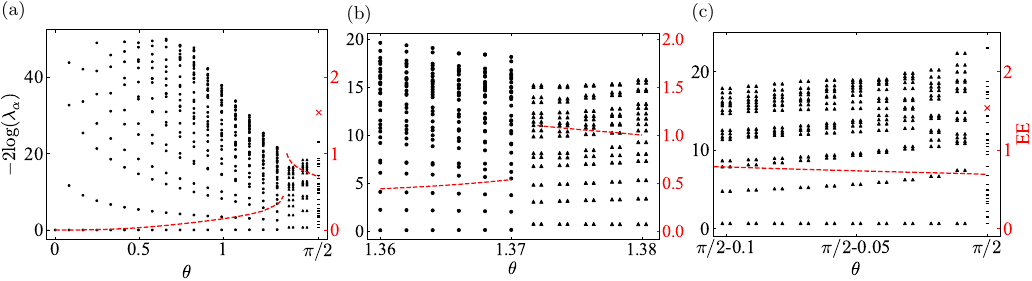}
\caption{Entanglement spectrum and EE of $\ket{\psi_{\text{fixed}}}$ for $\chi=32$. 
(a) $\theta\in[0,\pi/2]$. 
(b) $\theta\sim\theta_c$. 
(c) $\theta\sim\pi/2$.}
\label{Fig: Spectrum}
\end{figure*}

We use the VUMPS method to obtain the fixed-point iMPS $\ket{\psi_{\text{fixed}}}$ with $\chi=32$ for different $\theta$, then calculate the entanglement spectrum and the corresponding entanglement entropy (EE), as shown in Fig.~\ref{Fig: Spectrum}(a).
From the structure of the entanglement spectrum, we identify three phases throughout the range $\theta\in \left[0, \pi / 2\right]$.
We find that the entanglement spectrum of $\ket{\psi_{\text{fixed}}}$ is gapless at $\theta=\pi/2$ (i.e., measuring $X$ for all bulk qubits).
This result is consistent with a previous theoretical analysis~\cite{Liu2022}, where the authors have proved that $\theta = \pi/2$ shows a volume-law entanglement structure at the boundary, indicating we cannot reach a true fixed point as $L_y\rightarrow \infty$.
In Fig.~\ref{Fig: Spectrum}(b), we plot the entanglement spectrum of $\ket{\psi_{\text{fixed}}}$ near the $\theta = \pi/2$ point, which shows that any derivation from this point will bound the entanglement accumulation and $\ket{\psi_{\text{fixed}}}$ obeys the entanglement area law.

Furthermore, two phases exist in the area-law entanglement phase, whose entanglement spectra exhibit qualitative differences.
Through careful numerical simulations near the critical point in Fig.~\ref{Fig: Spectrum}(c), we find that the entanglement spectrum undergoes a qualitative change at $\theta_c = 1.371$.
For $\theta < \theta_c$, the entanglement spectrum is not degenerate, which is denoted as a trivial phase $\ket{\psi_{\rm trivial}}$ (in the sense of the entanglement structure).
For $\theta > \theta_c$, we observe that the entanglement spectrum of $\ket{\psi_{\text{fixed}}}$ is twofold degenerate, indicating a non-trivial entanglement structure $\ket{\psi_{\rm twofold}}$.
\subsection{Critical point} 
To clarify the type of phase transition between $\ket{\psi_{\rm trivial}}$ and $\ket{\psi_{\rm twofold}}$, we further study the ``first-excited'' state over the dominant eigenstate $\ket{\psi_{\rm fixed}}$.
Since the calculation of excitations under the VUMPS formalism strongly depends on the choice of ansatz, we adopt finite-size MPS with both an OBC and periodic boundary condition (PBC) to study the excited state.
Specifically, we first use the variational MPS method to calculate the dominant eigenstate $\ket{\psi_{0}}$ of $\hat{H}$ with energy $E_0$ for finite-size systems.
Then we calculate the dominant eigenvector of $\hat{H}-E_0\ket{\psi_0}\bra{\psi_0}$, which corresponds to the ``first-excited'' state of $\hat{H}$.

\begin{figure*}
\includegraphics[width=0.99\linewidth]{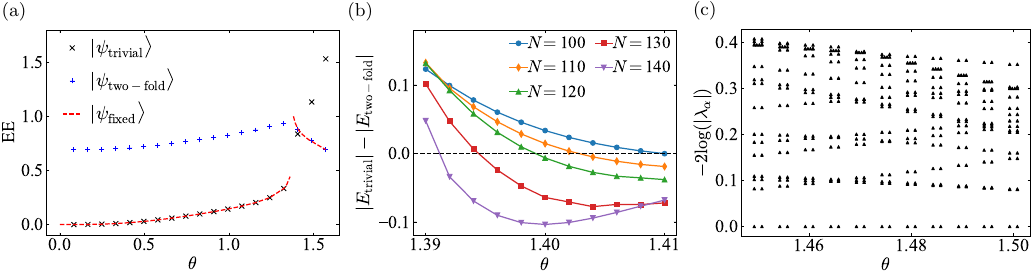}
\caption{(a) EE compared between OBC, $\chi=16$ and VUMPS, $\chi=32$.
(b) Energy gap (with PBC, $\chi=16$) for $\theta\sim \theta_c$.
(c) Moduli of the transfer matrix spectrum (with VUMPS, $\chi=32$) in the intermediate phase.}
\label{Fig: EE}
\end{figure*}

For systems with OBC, we calculate the entanglement spectra of these two states and find that one is consistent with the entanglement spectra of the trivial phase $\ket{\psi_{\rm trivial}}$, while the other is similar to that of the twofold degenerate phase $\ket{\psi_{\rm twofold}}$.
We also plot the EE of these two states throughout the range $[0, \pi/2]$ in Fig.~\ref{Fig: EE}(a), together with previous results obtained from the VUMPS method.
It is shown that the EE of $\ket{\psi_{\rm fixed}}$ calculated from the VUMPS method coincides with those of $\ket{\psi_{\rm trivial}}$ and $\ket{\psi_{\rm twofold}}$ in two phases, respectively, implying a shift of $\ket{\psi_{\rm fixed}}$ from $\ket{\psi_{\rm trivial}}$ to $\ket{\psi_{\rm twofold}}$ at the critical point.

This argument is further supported by the energy calculation for PBC, where we plot the energy gap between $\ket{\psi_{\rm trivial}}$ and $\ket{\psi_{\rm twofold}}$ in Fig.~\ref{Fig: EE}(b).
It is clearly demonstrated that there is a level crossing at the critical point for $N_s \geq 100$, revealing a first-order phase transition.

\subsection{Cat state} 
Here, we continue to study the entanglement structure of the state $\ket{\psi_{\text{fixed}}}$ in the intermediate phase whose entanglement spectrum is twofold degenerate.
We numerically demonstrate that the eigenvalues of the transfer matrix~\cite{PerezGarcia2007, Zeng2019} (or the double tensor, defined as $\mathbb{E}^{\alpha\gamma, \beta\chi} = \sum_i{T_{\ket{i}}^{\alpha\beta}T_{\ket{i}}^{\gamma\chi*}}$) appear in pairs $(\lambda, -\lambda)$, and their moduli are plotted in Fig.~\ref{Fig: EE}(c).

It has been proven in Ref.~\cite{PerezGarcia2007} that by adopting the canonical form, such a state can be further decomposed into the superposition of two states $\ket{\psi_{\rm fixed}} = \ket{\psi^0} + \ket{\psi^1}$, where both $\ket{\psi^0}$ and $\ket{\psi^1}$ are two-site periodic.
Here, $\ket{\psi^1} = \hat{T}\ket{\psi^0}$ and $\hat{T}$ is the translation operator for one site.

To fully explore this feature, we start with the points approaching $\theta\rightarrow \pi/2$, where the gap in the entanglement spectrum of $\ket{\psi_{\rm fixed}\left(\theta\rightarrow \pi/2\right)}$ diverges, as numerically demonstrated in Fig.~\ref{Fig: Spectrum}(c).
In other words, $\chi=2$ can provide an exact result in the limit $\theta\rightarrow \pi/2$.
In this limit, the transfer matrix $\mathbb{E}$ has four eigenvalues $\left\{\pm 1, \pm\frac{\sqrt{2}}{2}\right\}$.
In addition, the local tensor can be calculated as
\begin{align}
    T^{01} \rightarrow \ket{0},\, T^{10} \rightarrow \ket{+}.\label{equ: GHZ}
\end{align}
Therefore, the state can be explicitly reconstructed
\begin{align}
    \begin{aligned}
        \ket{\psi_{\rm fixed}{\left(\theta\rightarrow \pi/2\right)}} &= \cdots\otimes\ket{0}\otimes\ket{+}\otimes\ket{0}\otimes\ket{+}\otimes\cdots\\
        &+ \cdots\otimes\ket{+}\otimes\ket{0}\otimes\ket{+}\otimes\ket{0}\otimes\cdots\\
        &\equiv \ket{\psi^0}+\ket{\psi^1}.
    \end{aligned}
\end{align}
It can be directly verified that $\ket{\psi_{\rm fixed}{\left(\theta\rightarrow \pi/2\right)}}$ is an eigenvector of $\hat{H}$ with eigenvalue $E_0 = 1$.
Also, it is an eigenvector of the map from the virtual indices to the physical indices $\hat{O}$ at the upper boundary, i.e., $\ket{\psi_{\rm final}} = \hat{O}\ket{\psi_{\rm fixed}} = \ket{\psi_{\rm fixed}}$ for ${\theta\rightarrow \pi/2}$.
Obviously, both $\ket{\psi^0}$ and $\ket{\psi^1}$ are two-site periodic and $\ket{\psi^1} = \hat{T}\ket{\psi^0}$.
Meanwhile, these two components can also be connected by the parity operator, i.e., $\ket{\psi^1} = \hat{P}\ket{\psi^0}$, where $\hat{P}$ induces a spatial inversion regarding a bond.

As $\theta$ deviates from $\pi/2$, the entanglement gap becomes finite and higher-order Schmidt weights appear.
However, to study the global entanglement structure, it is sufficient to preserve only the Schmidt weights with the largest magnitude (and in this case we truncate $\ket{\psi_{\rm fixed}}$ to $\chi = 2$), since others only contribute to local entanglement and decay exponentially during the RG process.
It is numerically verified that, by properly choosing the gauge on virtual indices, the local tensor can be written as
\begin{align}
    T^{01} = \ket{\alpha},\, T^{10} = \ket{\beta},
\end{align}
where $\alpha$ and $\beta$ are general single-qubit states.
In other words, $\ket{\psi_{\rm fixed}}$ with $\chi=2$ can be decomposed as the superposition of two product states with each Schmidt weight corresponding to one state, i.e., 
\begin{align}
    \begin{aligned}
        \ket{\psi_{\rm fixed}}_{\chi=2} &= \cdots\otimes\ket{\alpha}\otimes\ket{\beta}\otimes\ket{\alpha}\otimes\ket{\beta}\otimes\cdots\\
        &+ \cdots\otimes\ket{\beta}\otimes\ket{\alpha}\otimes\ket{\beta}\otimes\ket{\alpha}\otimes\cdots\\
        &\equiv \ket{\psi^0}+\ket{\psi^1}.
    \end{aligned}
\end{align}
Therefore, these two components can be connected by either the translation operator $\hat{T}$ or the parity operator $\hat{P}$.

In summary, after removing local entanglement, the fixed-point iMPS $\ket{\psi_{\rm fixed}}$ in this twofold degenerate phase has two macroscopic components that can be connected to each other by either (a) translation of one site $\hat{T}$ or (b) parity operator regarding a bond $\hat{P}$.

\begin{figure*}
\centering
\includegraphics[width=0.99\linewidth]{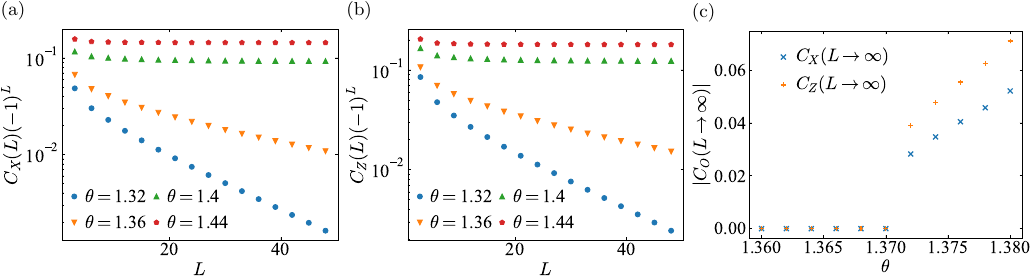}
\caption{
(a), (b) Two-site correlation function $C_X(L)$ and $C_Z(L)$ for $\ket{\psi_{\rm fixed}}$. 
(c) Convergence values for $|C_X(L\rightarrow\infty)|$ and $|C_Z(L\rightarrow\infty)|$ near the critical point $\theta_c$ for $\ket{\psi}_{\rm fixed}$.
The fixed-point iMPS are obtained from the VUMPS method with $D = 32$.}
    \label{correlation}
\end{figure*}

Since the iMPS in the twofold degenerate phase is non-injective, the correlation length defined as $\xi_x \equiv -1/\log{\left(|\lambda_2/\lambda_1|\right)}$ diverges throughout this phase, where $|\lambda_{1}|\geq|\lambda_{2}|$ are the eigenvalues of $\mathds{E}$ with the two largest magnitudes.
To detect this intermediate phase and the corresponding phase transition between $\ket{\psi_{\rm trivial}}$ and $\ket{\psi_{\rm twofold}}$, we evaluated the two-site correlation function $C_O(L)\equiv \braket{O_iO_{i+L}} - \braket{O_i}\braket{O_{i+L}}$ for $\ket{\psi_{\rm fixed}}$.
The results are shown in Figs.~\ref{correlation}(a) and \ref{correlation}(b) for $O = X$ and $O = Z$, respectively, where both correlators decay exponentially in the trivial phase.
On the contrary, they exhibit an oscillating behavior whose magnitudes converge to finite values in the twofold degenerate phase.
This phenomenon demonstrates the breakdown of translational symmetry and the existence of a long-range order in the fixed-point wave function.
The correlation functions for $L\rightarrow\infty$ are shown in Fig.~\ref{correlation}(c), exhibiting an abrupt drop at the critical point $\theta_c$, consistent with the previous statement of a first-order transition.
Therefore, these two-site correlation functions $C_X(L)$ and $C_Z(L)$ can serve as good detectors for this intermediate phase.

\begin{figure}
    \centering
    \includegraphics[width=0.96\linewidth]{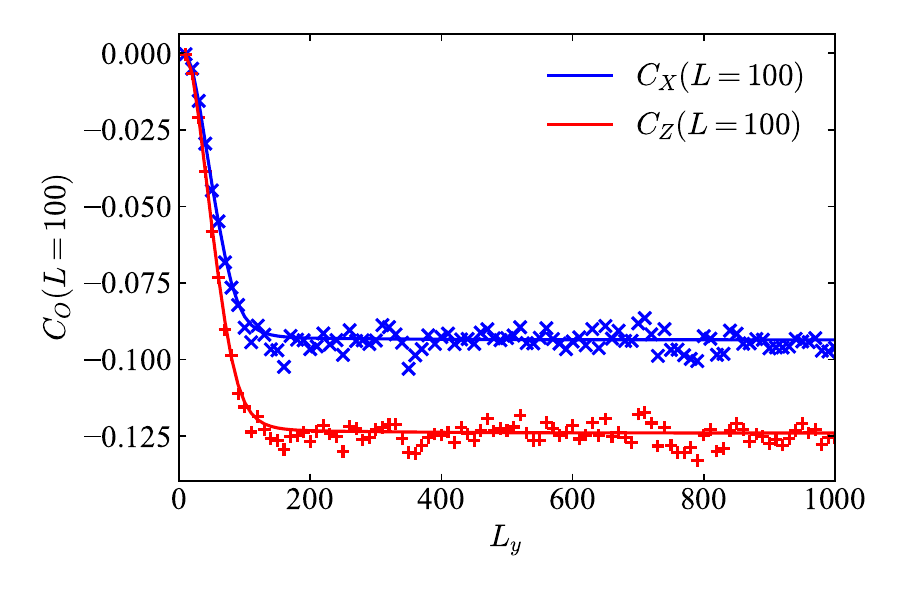}
    \caption{Dynamical evolution of correlation functions $C_X(L)$ and $C_Z(L)$ for $L = 100$ with and without measurement errors.
    We take $\theta = 1.4$ and the standard deviation of $\theta$ is $\varepsilon = 0.01$.}
    \label{Fig: Perturbation}
\end{figure}

To further study the inevitable errors when implementing the measurements in experiments, we assume that the measurement angle is randomly chosen from a Gaussian distribution whose standard deviation is $\varepsilon$ for each row in the system.
For simplicity, we still preserve the translational invariance within one row and simulate the dynamical properties with and without measurement noise.
In Fig.~\ref{Fig: Perturbation}, we simulate the dynamical system at $\theta = 1.4$, $\varepsilon=0.01$, and plot the correlation functions $C_X$ and $C_Z$ for $L = 100$ since any deviation from the condition $|\lambda_1| = |\lambda_2|$ will result in a final convergence of $C_O = 0$ for $L\rightarrow \infty$.
It is shown that the noisy system will converge to the twofold degenerate phase as soon as the ideal system does, after which the correlation functions will just fluctuate around the ideal values.
It means that the twofold degenerate phase at the boundary is robust against small errors in bulk measurements.

\subsection{Volume-law state at $\theta = \pi/2$}
Our VUMPS algorithm has difficulty in converging to a fixed-point iMPS at the $\theta = \pi/2$ point, where the entanglement spectrum is gapless and exhibits a volume-law behavior.
From the perspective of the 1+1D dynamical description of our system, i.e., $\ket{\psi_{\text{fixed}}}= \lim_{L_y\rightarrow \infty}\hat{H}^{L_y}\ket{\psi_{\text{init}}}$, it implies that $\hat{H}$ does not have a unique dominant eigenvector at $\theta=\pi/2$.
Analytically, it can be verified that $\hat{H}^{\dagger}\hat{H} = I$, where $\hat{H}$ involves an interaction with infinite length.
As a result, the eigenvalues $E$ of $H$ satisfy $|E| = 1$.
In other words, all the energy levels of $\hat{H}$, in the sense of magnitude, will collapse at $\theta = \pi/2$.
Therefore, we cannot reach a fixed point for $\lim_{L_y\rightarrow \infty}\hat{H}^{L_y}\ket{\psi_{\text{init}}}$, where all components will evolve simultaneously that enables the accumulation of entanglement, resulting in a volume-law phase at the boundary.
This argument can also explain why the entanglement spectrum of $\ket{\psi_{\rm fixed}}$ changes abruptly just at the point $\theta = \pi / 2$ since any derivation from this point will lead to a final convergence to $\ket{\psi_{\rm twofold}}$, which satisfies the area-law entanglement.

\begin{figure*}
    \centering
    \includegraphics[width=0.8\linewidth]{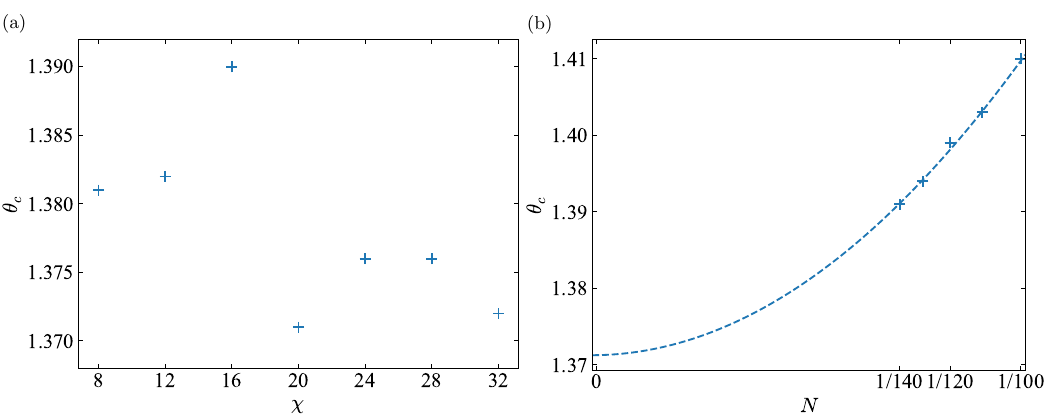}
    \caption{Critical points $\theta_c$ for (a) different $\chi$ with VUMPS and (b) different $N$ with PBC and $\chi=16$.}
    \label{Fig: Critical}
\end{figure*}

\section{Conclusions and Discussions}
In this paper we investigate a measurement-based entanglement phase transition on the boundary state of an infinite-size two-dimensional cluster state. 
The state is subjected to uniform measurements $M = \cos{\theta}Z+\sin{\theta}X$ on the lower boundary qubits and in all bulk qubits.
Our results show that at $\theta = \pi/2$, the system exhibits a volume-law entanglement phase, consistent with previous studies~\cite{Liu2022}.
Conversely, for any $\theta<\pi/2$, the system converges to a fixed-point iMPS $\ket{\psi_{\rm fixed}}$ with area-law entanglement.

Moreover, we provide further insight into the system by identifying two phases within the area-law region and a level-crossing phase transition between them at $\theta_c=1.371$. 
For $\theta<\theta_c$, the entanglement spectrum is trivial, and the boundary state is short-range correlated, which can be smoothly connected to a product state. 
However, in the second phase, we observe a non-injective iMPS $\ket{\psi_{\rm fixed}}$ exhibiting long-range correlations and two macroscopic components. 
These components are both two-site periodic and can be related to each other by the translation of a single site. 
Furthermore, the two components are related by the parity operator $\hat{P}$ up to local unitary transformations.

It is worth noting that this phase cannot be realized as the unique ground state of a 1D local, gapped Hamiltonian.
This demonstrates that the boundary of a two-dimensional system can exhibit a phase diagram more complex than that of a standard one-dimensional system, which is the fundamental idea behind MBQC.
It also indicates that the topological properties of bulk PEPS with appropriate measurements can be reflected in edge degrees of freedom~\cite{Cirac2011, Schuch2013, Yang2014, Yang2015}.
Additionally, this phase is robust under small perturbations, where the measurement angle deviates slightly from its preset value.
However, we must point out that the phase diagram in our study requires postselection, which means it is difficult to be observed experimentally at the current stage.

An interesting direction for future research would be to extend our results to cases with nonuniform measurements on each site.
For instance, if we maintain translational invariance along the $x$ direction but allow for a varying pattern of $\theta$ along the $y$ direction, we will encounter a relaxation process.
If the characteristic length of the varying $\theta$ is much larger than the convergence length $\xi_y$ of the system, where $\xi_y\sim 1/ \Delta$, with $\Delta$ being the energy gap between $\ket{\psi_{\rm trivial}}$ and $\ket{\psi_{\rm twofold}}$, the system will adiabatically evolve between $\ket{\psi_{\rm fixed}}$ states with different $\theta$.
However, if $\theta$ changes rapidly along the $y$ direction, we can only capture the average effect for large $L_y$.
The intermediate region will be much more complex and interesting, where we may discover possible dynamical phase transitions for specific patterns of $\theta$, which we leave for future study.

\begin{acknowledgments}
We thank Timothy H. Hsieh for insightful discussions. Y.G. and S.Y. are supported by the National Natural Science Foundation of China (NSFC) (Grants No. 12174214 and No. 92065205), the National Key R\&D Program of China (Grant No. 2018YFA0306504), the Innovation Program for Quantum Science and Technology (Grant No. 2021ZD0302100), and the Tsinghua University Initiative Scientific Research Program. J.H.Z. and Z.B. are supported by a startup fund from the Pennsylvania State University and thankful for the hospitality of the Kavli Institute for Theoretical Physics, which is partially supported by the National Science Foundation under Grant No. NSF PHY-1748958.
\end{acknowledgments}

\appendix
\section{Finite bond dimension and finite-size scaling}
In Fig.~\ref{Fig: Critical} (a), critical points $\theta_c$ calculated under different bond dimensions $\chi$ with the VUMPS method are plotted, where $\theta_c$ vary with $\chi$ in the region $[1.37, 1.39]$.
It is expected that $\theta_c$ will converge to a finite value smaller than $\pi/2$ for $\chi\rightarrow\infty$, ensuring the existence of the intermediate phase.
At the same time, we also explore the finite-size effect of our PBC calculation in Fig.~\ref{Fig: Critical} (b), where the critical points are visually inspected from the results in Fig.~\ref{Fig: EE} (b).
With a quadratic fitting, we extrapolate the critical point for $N\rightarrow +\infty$ and obtain an estimated value $\theta_c = 1.371$, consistent with the results from VUMPS for the infinite system.

\bibliography{ref}

\end{document}